# Dual Flat-Bands of Bound State in the Continuum and Radiative Mode via TE-TM Coupling

Jiayao Liu, Zimeng Zeng, Zhuoyang Li, Zelong He, Zhaona Wang*

Key Laboratory of Multiscale Spin Physics (Ministry of Education), Applied Optics Beijing Area Major Laboratory, School of Physics and Astronomy, Beijing Normal University, Beijing 100875, China

*Contact author: zhnwang@bnu.edu.cn*

## Abstract

A general symmetry-controlled mechanism is proposed for realizing dual flat-bands of bound state in the continuum (BIC) and its radiative counterpart in photonic crystal slabs. By breaking the vertical mirror symmetry of slab, inter-polarization coupling between TE-like and TM-like modes is activated, while intra-polarization coupling among modes within the same polarization class is simultaneously preserved. The cooperative action of these two coupling channels gives rise to the concurrent flattening of both the BIC-hosting band and the radiative band, resulting in a dual flat-band system with strongly contrasting quality ($Q$) factors. An effective two-step coupling model is constructed to capture the essential physics and show that the emergence of the flat bands is governed by geometric tuning rather than accidental degeneracies. The mechanism is shown to be generic with respect to polarization and material platform, enabling dual flat-band states in both low- and high-index systems, with substantially enhanced angular bandwidths in the latter. These finding establish a unified route for flat-band photonic engineering and provide a robust platform for angle-tolerant resonant photonic functionalities.

# I. INTRODUCTION

Photonic flat-band modes characterized with dispersionless nature and vanishing group velocity provide a powerful platform for enhancing light-matter interactions through a high density of states (DOS) and robust wide-angle responses [1-5]. These properties have stimulated extensive research into low-threshold lasing [6,7], nonlinear optics [8], and polariton condensation [9,10]. Meanwhile, bound states in the continuum (BICs), first proposed in Quantum mechanics by Wigner and von Neuman [11,12], offer another paradigm of light confinement, achieving theoretically infinite quality factors through destructive interference in the radiation continuum. In period photonic crystal (PC), BICs are often associated with nontrivial topological charges, manifesting as polarization singularities in the far field [13,14]. Various BIC-related physical phenomena have been reported, including symmetry-breaking-induced quasi-BIC [15], merging BIC [16,17], chiral BIC [18,19], exceptional BIC [20], and Janus BIC [21]. Despite these advances, conventional photonic BICs typically occur at isolated points in momentum space and reside on dispersive Bloch bands. Consequently, symmetry breaking induced by oblique incidence drives the BIC into a quasi-BIC regime, accompanied by pronounced frequency shifts and a rapid degradation of the quality factor, which severely limits their performance in wide-angle applications [22]. This inherent limitation motivates the integration of flat-band dispersion with BIC physics, creating flat-band BICs that simultaneously suppressed radiation loss and exhibit vanishing group velocity. Such modes promise to deliver ultra-high-$Q$ resonances that remain robust over a broad angular range, which are highly desirable for advanced sensing[23], lasing[24,25], and nonlinear photonics[26,27].

To date, several strategies have been proposed to realize flat-band BICs, including moiré photonic lattices [22,27,28], band-folding technique [26,29-32], and unit-cell geometry engineering [23-25,33-36]. However, two fundamental limitations persist. First, nearly all realizations rely on high-index dielectric platforms (e.g. silicon or $TiO_2$), where strong mode confinement is readily available. The extension of flat-band BIC concepts to low-index systems remains comparatively unexplored, primarily due to their inherently weaker light-field confinement, despite their practical advantages in fabrication compatibility and reduced material losses [37]. Second, existing designs almost exclusively exploit intra-coupling between modes of the same polarization class of transverse electric (TE) or transverse magnetic (TM) modes. The possibility of generating flat bands through direct inter-polarization coupling between TE and TM modes has not been systematically addressed. This omission is significant, because polarization represents a fundamental, yet underexploited, degree of freedom in photonic band engineering.

Furthermore, a BIC is intrinsically accompanied by a radiative partner mode of much lower $Q$ factor in the two Bloch mode coupling systems. In most conventional flat-band BIC designs, this radiative mode is often treated as a mere by-product. For example, Le *et al.* [34] realized a flat-band BIC in a one-dimensional (1D) silicon PC slab via symmetry-breaking-induced coupling between even and odd modes of same polarization. In two-dimensional (2D) $TiO_2$ PC slab, Do *et al*. [24] achieved a flat-band BIC by tuning the lattice period to control the coupling among multiple modes with identical polarization. In both cases, however, the bands of radiative counterparts remain dispersive. A compelling but unanswered question emerges: Can both the BIC and its radiative partner be simultaneously engineered into flat-band modes, creating a dual flat-band system that combines ultra-high $Q$ and efficient radiation within the same platform?

Here, we introduce a universal symmetry-control strategy that resolves these limitations simultaneously. By deliberately breaking vertical mirror symmetry in a waveguide-PC-coupled photonic slab, direct coupling between TE-like and TM-like modes is enabled. Combined with carefully engineered intra-polarization coupling among modes within the same polarization class, this symmetry-breaking-induced inter-polarization coupling enables the concurrent formation of a flat-band BIC and its radiative counterpart, resulting in a dual flat-band system with distinctly different $Q$ factors. Crucially, the underlying symmetry-control principle generates the dual flat-band system in both high- and low-index platforms, bridging a longstanding divide in photonic design. Our work establishes inter-polarization coupling as a new and powerful knob for concurrent flat-band engineering. It delivers a unified platform that combines angle-robust high-$Q$ resonance with efficient radiation, opening routes to multifunctional meta-devices that can be implemented across a wide range of material systems.

## II. THEORETICAL MODEL AND CALCULATION RESULTS

*Theoretical model.* In a 2D PC slab with vertical mirror-flip symmetry ($z \rightarrow -z$), as schematically illustrated in Fig.1(a). The structure is periodic along $x$ and $y$ directions with lattice constants $P_x$ and $P_y$ ($P_x < P_y$) and consists of circular air holes with diameter $d$ patterned in a dielectric slab of finite thickness $h_1$ along the $z$ direction. Owing to the vertical mirror symmetry of the structure, all resonant modes can be classified into TM-like ($H_x$, $H_y$, 0) and TE-like ($E_x$, $E_y$, 0) modes[16,38]. TM-like modes are symmetry-decoupled from TE-like modes, and mode interactions occur only within the same polarization class. As shown in Fig. 1(b), the TM-like modes (red lines) remain uncoupled from the TE-like modes (black lines), and coupling arises only among the modes with identical polarization. The TM-like bands originate from four Bloch resonances ($\pm 1$, 0)$_{TM}$ and (0, $\pm 1$)$_{TM}$, whereas the TE-like bands are derived from two Bloch resonances ($\pm 1$, 0)$_{TE}$. Here, the integers $m$ and $n$ denote the diffraction orders along the $x$ and $y$ directions, respectively, and the corresponding Bloch wave vectors are given by[24,35,39,40]

$$\boldsymbol{\beta}_{m,n} = (k_x + m\frac{2\pi}{P_x})\boldsymbol{u}_x + (k_y + n\frac{2\pi}{P_y})\boldsymbol{u}_y. \qquad (1)$$

The intra-polarization coupling among the four Bloch resonance can be described by a 4×4 Hamiltonian[24,39,41]

$$H_{\text{intra}} = \begin{bmatrix} \omega_1^\alpha & U_1^\alpha & U_3^\alpha & U_3^\alpha \\ U_1^\alpha & \omega_2^\alpha & U_3^\alpha & U_3^\alpha \\ U_3^\alpha & U_3^\alpha & \omega_3^\alpha & U_2^\alpha \\ U_3^\alpha & U_3^\alpha & U_2^\alpha & \omega_4^\alpha \end{bmatrix} + i \begin{bmatrix} \gamma_1^\alpha & \Gamma_{12}^\alpha & \Gamma_{13}^\alpha & \Gamma_{14}^\alpha \\ \Gamma_{12}^\alpha & \gamma_2^\alpha & \Gamma_{23}^\alpha & \Gamma_{24}^\alpha \\ \Gamma_{13}^\alpha & \Gamma_{23}^\alpha & \gamma_3^\alpha & \Gamma_{34}^\alpha \\ \Gamma_{14}^\alpha & \Gamma_{24}^\alpha & \Gamma_{34}^\alpha & \gamma_4^\alpha \end{bmatrix}, \qquad (2)$$

where $\alpha$ = TM or TE, $\omega_i^\alpha$ and $\gamma_i^\alpha$ ($i = 1, \ldots, 4$) are the real and imaginary parts of the uncoupled Bloch resonances modes ($\pm1, 0$)$_{TM/TE}$ and ($0, \pm1$)$_{TM/TE}$, respectively. $U_1^\alpha$ and $U_2^\alpha$ are coupling strength between counter-propagating modes along the $x$ and $y$ direction, respectively, while $U_3^\alpha$ is coupling strength between modes propagating orthogonal directions. The imaginary off-diagonal terms are given by $\Gamma_{i,j} = \sqrt{\gamma_i \gamma_j}(\boldsymbol{u}_i^\alpha \cdot \boldsymbol{u}_j^\alpha)$, $\boldsymbol{u}_i^\alpha$ denotes the far-field polarization unit vectors of the corresponding uncoupled mode, given by[24,34]

$$\boldsymbol{u}^{TM} = \frac{1}{\sqrt{(m/P_x + k_x)^2 + (n/P_y + k_y)^2}} \cdot (m/P_x + k_x, n/P_y + k_y), \qquad (3)$$

and

$$\boldsymbol{u}^{TE} = \frac{1}{\sqrt{(m/P_x + k_x)^2 + (n/P_y + k_y)^2}} (-(n/P_y + k_y), m/P_x + k_x). \qquad (4)$$

In this coupling regime, a symmetry-protected BIC and its radiative counterpart emerge at the $\Gamma$ point from the coupling of the TM-like Bloch resonances $(0, \pm1)_{TM}$. Increasing the air-hole diameter $d$ enhances the intra-polarization coupling strength among TM-like modes, leading to the formation of a flat-band BIC, as shown in Fig. 1(c). In contrast, the band associated with the radiative mode remains strongly dispersive. This behavior reflects a fundamental limitation of vertically symmetric structures: while a flat-band BIC can be realized through intra-polarization coupling, the radiative mode lacks additional coupling channels required for band flattening. As a result, the simultaneous realization of a flat-band radiative mode is prohibited by symmetry.

To overcome this constraint, vertical symmetry breaking in a waveguide-PC-coupled configuration is introduced by reducing the PC thickness to $h_2 < h_1$, as illustrated in Fig. 1(d). This symmetry-breaking activates inter-polarization coupling between TE-like and TM-like radiative modes. The resulting inter-polarization interaction can be described by a 2×2 Hamiltonian[34]

$$H_{inter} = \begin{bmatrix} \omega_y^{TM} & \kappa \\ \kappa & \omega_x^{TE} \end{bmatrix} + i \begin{bmatrix} \gamma_y^{TM} & \sqrt{\gamma_y^{TM} \cdot \gamma_x^{TE}} e^{-i\phi} \\ \sqrt{\gamma_y^{TM} \cdot \gamma_x^{TE}} e^{i\phi} & \gamma_x^{TE} \end{bmatrix}, \quad (5)$$

where $\omega_y^{TM}$ and $\omega_x^{TE}$ denote the resonance frequencies, $\gamma_y^{TM}$ and $\gamma_x^{TE}$ represent the corresponding radiative losses. The phase $\phi$ is the relative dephasing when the two modes couple to a common radiative channel, while $\kappa$ is coupling strength and can be approximated as $\kappa = \kappa_0 |k_x P_x / 2\pi|$ [34]. For appropriately structural parameters, this additional coupling channel enables the concurrent formation of a flat-band BIC and a flat-band radiative mode, thereby realizing a dual flat-band system, as shown in Fig. 1(e).

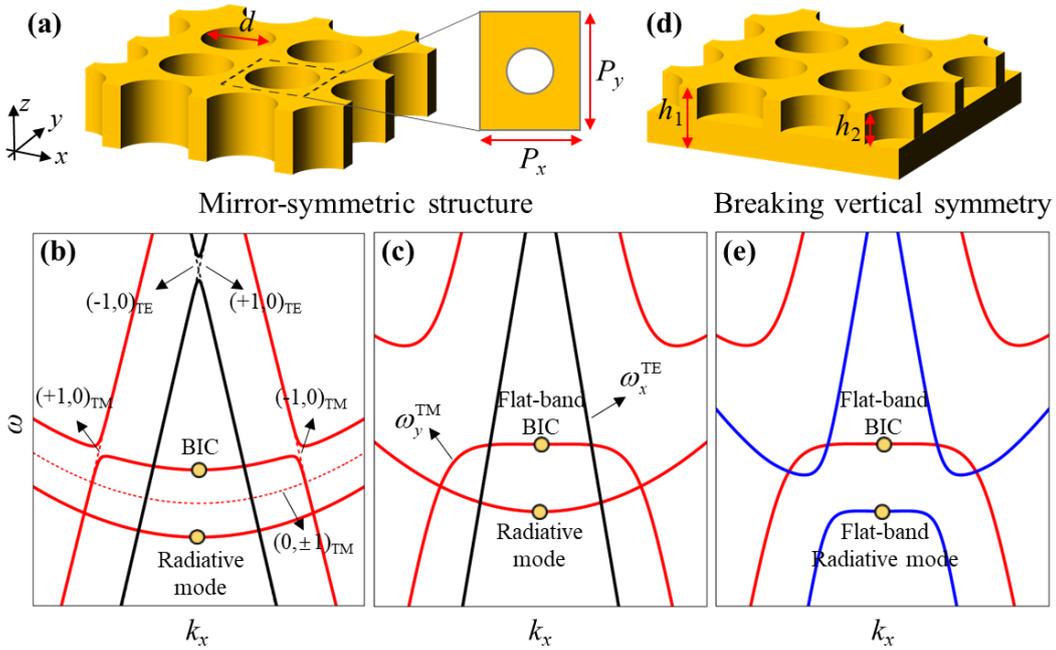

FIG. 1. Mechanism for realizing dual flat-band modes. (a) Schematic of the PC slab possessing mirror symmetry. (b, c) Band dispersions along the $k_x$ direction for the vertically symmetric structure, showing a dispersive band hosting a BIC(b) and a flat-band BIC (c), respectively. (d) Schematic of the PC slab with broken mirror symmetry. (e) Band dispersion along the $k_x$ direction for the symmetry-broken structure, where a flat-band BIC emerges simultaneously with a flat-band radiative mode.

*Engineering dispersion in real structure.* As an example, the realization of TM-like dual flat-bands in a real structure is considered. The slab has a refractive index $n = 1.65$, lattice constants are $P_x = 330$ nm and $P_y = 380$ nm, thickness $h_1 = 200$ nm, and air-hole diameter $d = wP_y = 0.2P_y$. The corresponding band structure of the symmetric PC slab, calculated using the finite-element method (FEM), is shown in Fig. 2(a), with the color scale representing the $Q$ factor. For clarity, only three TM-like bands ($TM_1$, $TM_2$, $TM_3$) and one TE-like band ($TE_1$) within the frequency range of interest are displayed. A symmetry-protected BIC appears at the $\Gamma$ point of $TM_2$ band, accompanied by a radiative mode counterpart with a much lower $Q$ factor on the adjacent band $TM_3$[24,42]. Away from the $\Gamma$ point, the BIC hosting-band $TM_2$ exhibits hybridization with nearby modes, while the radiative mode band $TM_3$ and $TE_1$ remain largely isolated.

The vertical mirror symmetry is then broken by reducing the etch depth to $h_2 = 100$ nm, while keeping the air-hole diameter unchanged. This symmetry breaking enables inter-polarization coupling between the radiative mode band $TM_3$ and TE-like band $TE_1$, leading to pronounced band hybridization (labeled as I and II in the Fig. 2(b)). By further increasing the air-hole diameter to $d = 0.4P_y$, the inter-polarization coupling strength is significantly enhanced. As a result, both the BIC-hosting band III and the radiative mode band II progressively flatten, giving rise to the simultaneous formation of a flat-band BIC and a flat-band radiative mode, as shown in Fig. 2(c). The corresponding magnetic-field profiles of the two flat-band modes are presented in Figs. 2(d) and 2(e). The BIC exhibits an antisymmetric field distribution, which suppresses radiation into the far field, whereas the radiative mode displays a symmetric field profile, allowing efficient coupling to free space radiation[43].

The far-field polarization properties of the flat-band BIC on band III are further analyzed in the vicinity of the $\Gamma$ point. The far-field polarization vector $(C_x, C_y)$ is obtained by projecting the far-field polarization onto *x-y* plane[16,17,44], where $(C_x, C_y) = \iint e^{ik_x x + ik_y y} \boldsymbol{E}(k_x, k_y) dx dy \Big/ \iint dx dy$. To characterize the polarization state, the polarization orientation angle $\phi = 1/2 \arg(S_1 + iS_2)$ and the ellipticity angle

$\chi = 1/2 \arctan(S_3 / \sqrt{S_1^2 + S_2^2})$ are introduced, where the Stokes parameters are given by $S_0 = |C_x|^2 + |C_y|^2$, $S_1 = |C_x|^2 - |C_y|^2$, $S_2 = 2\operatorname{Re}(C_x C_y^*)$, and $S_3 = 2\operatorname{Im}(C_x C_y^*)$. The calculated far-field polarization texture is shown in Fig. 2(f), where the BIC is identified and marked by a polarization vortex, corresponding to a singular point of the polarization field. The topological charge $q$ associated with the BIC is evaluated from the winding number of the polarization vector along a closed loop $C$ encircling the BIC in momentum space, defined as $q = \frac{1}{2\pi} \oint_C dk \cdot \nabla_k \phi(k)$. Along the counter-clockwise loop, the polarization vector undergoes a full $+2\pi$ rotation, yielding a topological charge of $q = +1$ for the flat-band BIC[24,45]. These results demonstrate that the dual flat-bands consisting of a BIC and its radiative counterpart can be simultaneously realized in a low-index photonic platform via the combined effects of symmetry breaking and coupling-strength engineering.

The band dispersion is further analyzed by two-step model based on Eqs. (2) and (5), which enables a systematic examination of the flat-band evolution in the system. The analysis is performed within a complete basis of eight uncoupled modes, consisting four TM-like modes, $(\pm 1, 0)_{TM}$ and $(0, \pm 1)_{TM}$, and four TE-like modes, $(\pm 1, 0)_{TE}$ and $(0, \pm 1)_{TE}$ (see Appendix A for details). In the vicinity of the $\Gamma$ point, the dispersion relation of each uncoupled mode is approximated by a linear expansion derived from Eq. (1) [24]

$$\omega_{m,n} = \omega_0 + \frac{c}{n_g}(\sqrt{(k_x + m\frac{2\pi}{P_x})^2 + (k_y + n\frac{2\pi}{P_y})^2} - \sqrt{(m\frac{2\pi}{P_x})^2 + (n\frac{2\pi}{P_y})^2}), \quad (6)$$

where $c$ is the speed of light in vacuum, $n_g$ is group refractive index, and $\omega_0$ denotes the angular frequency of the uncoupled mode $(m, n)$ at the $\Gamma$ point. For the lowest-order TM-like and TE-like modes, $\omega_0$ can be determined from the slab waveguide dispersion relations[46-48]

$$h_1\sqrt{(\frac{\omega_0^{TM}}{c})^2 n_{slab}^2 - A} = 2\arctan\left(\frac{n_{slab}^2\sqrt{A-(\frac{\omega_0^{TM}}{c})^2 n_0^2}}{n_0^2\sqrt{(\frac{\omega_0^{TM}}{c})^2 n_{slab}^2 - A}}\right), \tag{7}$$

$$h_1\sqrt{(\frac{\omega_0^{TE}}{c})^2 n_{slab}^2 - A} = 2\arctan\left(\frac{\sqrt{A-(\frac{\omega_0^{TE}}{c})^2 n_0^2}}{\sqrt{(\frac{\omega_0^{TE}}{c})^2 n_{slab}^2 - A}}\right), \tag{8}$$

where $A = (m\frac{2\pi}{P_x})^2 + (n\frac{2\pi}{P_y})^2$, $n_{slab}$ and $n_0$ are the effective refractive indices of the slab and surrounding medium, respectively.

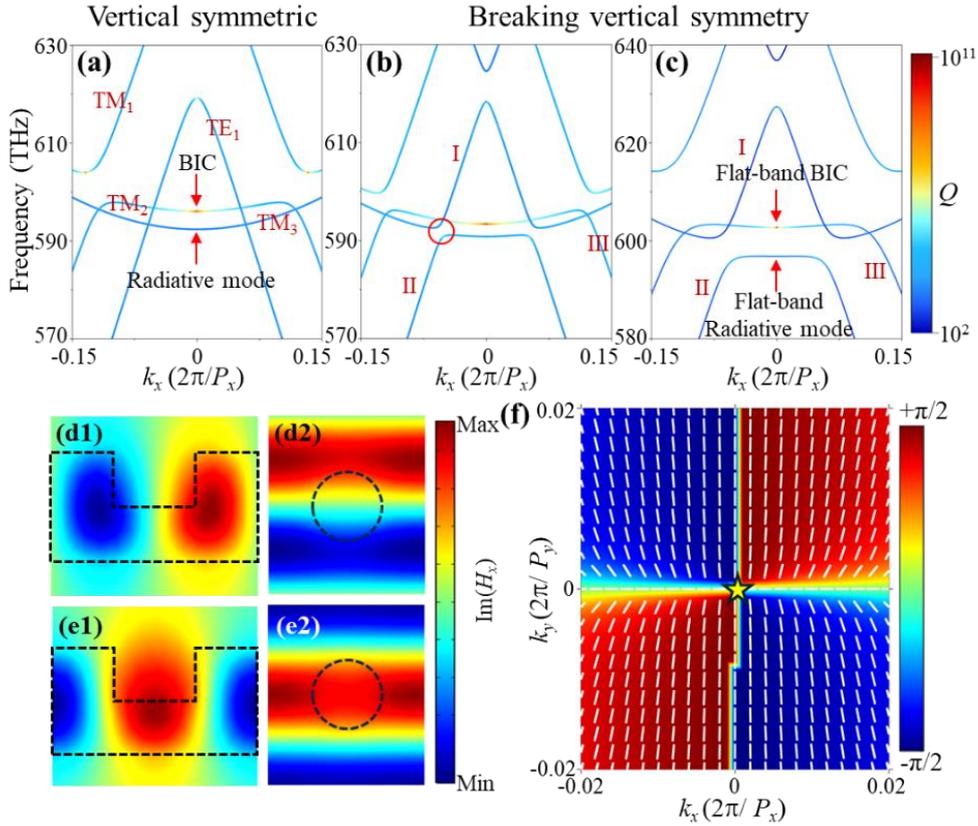

FIG. 2. (a) Calculated photonic band structure of vertically symmetric slab, where the color scale represents the $Q$-factor. (b, c) Calculated band structure of the vertically symmetry-broken slab with air-hole diameters $d = 0.2P_y$ (b) and $d = 0.4P_y$ (c), respectively. (d1, d2) Magnetic-field distributions of the BIC at the Γ point of band III

in the *y-z* and *x-y* planes, respectively. (e1, e2) Magnetic-field distributions of radiative mode at the Γ point of band II in the *y-z* and *x-y* planes, respectively. (f) Simulated far-field polarization texture of the BIC mode near the Γ point.

Figure 3(a) compares the angle-resolved reflection spectra obtained from rigorous coupled-wave analysis (RCWA)[49] simulations with the band dispersions predicted by the effective Hamiltonian. All parameters entering the two-step model are uniquely extracted from the RCWA results and summarized in Appendix A. The excellent agreement between the numerical spectra and the analytical dispersions confirms the validity of the effective model. Based on this model, the evolution of the flat bands is systematically investigated by exploring the structural parameter space. Specifically, the lattice period $P_x$ and the slab thickness $h_1$ are varied, while all coupling coefficients are kept fixed. For each parameter set, the angular width $\Delta q = \Delta k_x P_x / 2\pi$ of the flat-band BIC and flat-band radiative mode is evaluated under the flatness criteria $\Delta \hat{\omega} = \omega P_y / 2\pi c \leq 7 \times 10^{-4}$ and $d\hat{\omega}/dq \leq 10^{-2}$ within the considered angular range. The resulting angular widths for the flat-band BIC and the flat-band radiative mode are shown in Figs. 3(b) and 3(c), respectively. As $P_x$ and $h_1$ are varied, the angular extents of both flat bands can be continuously tuned. Three representative parameter regimes are identified: the coexistence of a flat-band BIC and a flat-band radiative mode (blue pentagram), a flat-band BIC only (black pentagram), and a flat-band radiative mode only (red pentagram). The corresponding band dispersions are shown in Figs. 3(d)-3(f). These results demonstrate that the emergence of dual flat-band modes arises from a controlled interplay between lattice periodicity and slab geometry, rather than from accidental degeneracies, highlighting the critical role of geometric tuning in flat-band engineering.

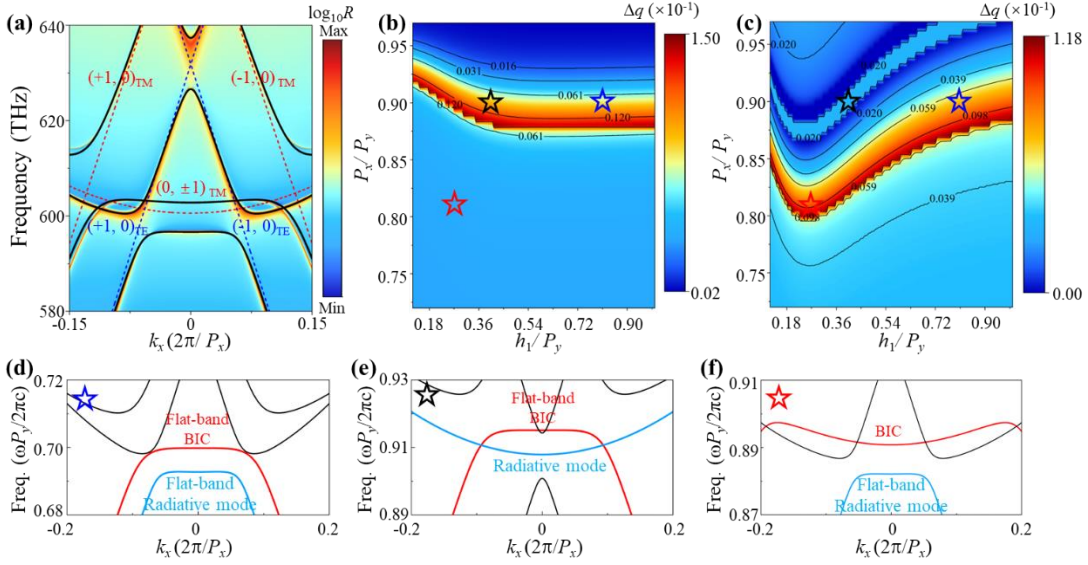

FIG. 3. (a) Comparison between the band structures obtained from RCWA simulations and those calculated using the effective Hamiltonian model. The black lines represent the bands from the Hamiltonian, while the red and blue lines denote the uncoupled TM-like and TE-like guided modes, respectively. (b, c) Angular widths $\Delta q$ of the flat-band BIC (b) and the flat-band radiative mode (c) as functions of the lattice period $P_x$ and slab thickness $h_1$, respectively. (d-f) Band structures corresponding to the marked parameter sets in (b) and (c) which are coexistence of the flat-band BIC an a flat-band radiative mode (d), only the flat-band BIC (e), and only the flat-band radiative mode (f), respectively.

Additionally, distinctly polarization-dependent spectral responses are observed between the two TM-like flat-band modes under s- and p-polarized excitations (see Appendix B for details). Specifically, s-polarized incidence corresponds to an electric field polarized along the $y$ direction, whereas p-polarized incidence corresponds to a magnetic field polarized along the same direction. Under s-polarized excitation, the flat-band radiative mode is efficiently excited, while under p-polarized excitation, the flat-band BIC is selectively accessed. This pronounced polarization selectivity enables controllable switching between low-$Q$ and high-$Q$ flat-band states within the same photonic platform.

To demonstrate the generality of the proposed mechanism for realizing dual flat-band modes, the analysis is extended to the TE-like polarization, as shown in Fig. 4. The slab band structure in Fig. 4(a) exhibits two nearly dispersionless modes at the Γ point, corresponding to a flat-band BIC and a flat-band radiative mode. The angular bandwidths reach 10.2° (±5.1°) for flat-band BIC and 11.8° (±5.9°) for the flat-band radiative mode, while the relative wavelength variation remains below Δλ/λ < 0.026% within the flat-band angular range. The associated field distributions, presented in Figs. 4(b) and 4(c), indicate that the flat-band BIC features an antisymmetric field profile, whereas the flat-band radiative mode displays a symmetric field distribution. The BIC character is further confirmed by a singular polarization vortex at the Γ point and by the momentum-resolved $Q$-factor distribution in Fig. 4(d). These results demonstrate that the proposed mechanism is polarization-independent, supporting both TM-like and TE-like dual flat bands within the same photonic platform.

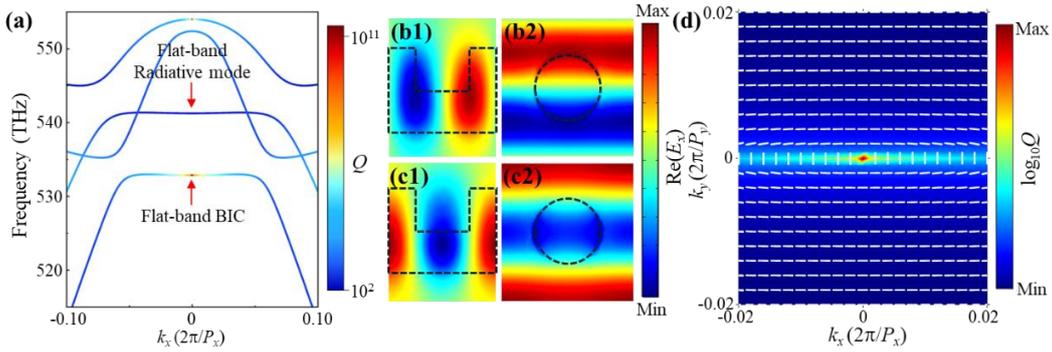

FIG. 4. (a) Calculated band structure of the PC slab with $h_1 = 2h_2 = 300$ nm, $d = 0.5P_y = 380$ nm, and $P_x = 360$ nm. The color scale indicates the $Q$ factor. (b1, b2) Electric-field distributions of the BIC at the Γ point in the $y$-$z$ and $x$-$y$ planes, respectively. (c1, c2) Electric-field distributions of the radiative mode at the Γ point in the $y$-$z$ and $x$-$y$ planes, respectively. (d) Simulated far-field polarization texture and corresponding $Q$-factor distribution of the BIC near the Γ point.

Importantly, the proposed approach exhibits strong material versatility and remains applicable over a broad range of refractive indices. Dual TE-like flat bands are demonstrated in a high-index silicon PC slab with $n = 3.41$. As shown in Fig. 5(a), two nearly dispersionless bands corresponding to a BIC and a radiative mode emerge near the Γ point. The corresponding excitation angular bandwidths Δθ of dual flat bands are significantly expanded, reaching 21.4° (±10.7°) for BIC band and 32.2° (±16.1°) for radiation mode, respectively. This large angular bandwidth is from the synergistic effect of the intra-polarization and inter-polarization coupling. Notably, the relative wavelength variation remains below $\Delta\lambda/\lambda < 0.026\%$, indicating a ultra-flatness band compared to that reported for silicon dielectric pillar arrays in Ref. [50]. To clarify that the band-dispersion control originates from structural engineering rather than material-specific parameters, the dual flat bands are further investigated in PC slabs composed of materials spanning from low to high refractive indices. The excitation angular bandwidths summarized in Fig. 5(b) show that both the flat-band BIC and the flat-band radiative mode persist across a broad refractive-index range, while systematically broadening with increasing refractive index. These results establish that the proposed dual flat-band mechanism is not limited to a particular material platform and can be readily scaled to high-index photonic systems, offering enhanced angular tolerance and improved prospects for integrated photonic applications.

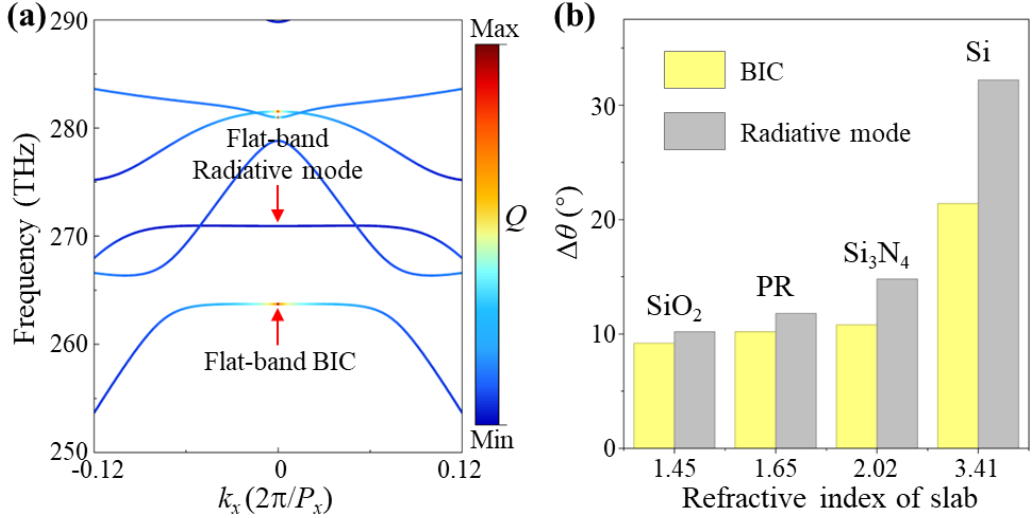

FIG. 5. (a) Calculated band structure of high-index PC slab with $P_x$ = 355 nm and refractive index $n$ = 3.41. The color scale represents the $Q$ factor. (b) Excitation angular bandwidths $\Delta\theta$ of the dual flat-bands for differently material platforms, including $SiO_2$, photoresist (PR), $Si_3N_4$, and Si.

## III. CONCLUSIONS

In summary, we have proposed and demonstrated a general mechanism for realizing dual flat-band modes consisting of a flat-band BIC and a flat-band radiative mode in PC slabs. By jointly engineering intra-polarization coupling among modes of the same polarization class (TM-like or TE-like) and symmetry-breaking-induced inter-polarization coupling between TM-like and TE-like modes in a symmetry-breaking waveguided-grating-slab, both the BIC-hosting band and its radiative counterpart can be simultaneously flattened. This coupling mechanism provides a clear physical picture for the emergence of dual flat-bands and is accurately captured by an effective Hamiltonian model. The proposed strategy is shown to be independent of both polarization and material parameters. Dual flat-bands are realized for both TM-like and TE-like modes and persist across a wide refractive-index range, from low-index photonic platforms to in high-index systems. In particular, in silicon PC slab, the angular bandwidth of flat-band BIC and radiative mode reach ±10.7° and ±16.1°, respectively, while maintain an ultrahigh flatness characterized by a relative wavelength variation remains below $\Delta\lambda/\lambda < 0.026\%$. Owing to their distinct radiation properties, the two flat-band modes can be selectively excited by external fields with different polarizations (s- or p-polarized), enabling flexible access to high-$Q$ and low-$Q$ resonant responses within the same structure. These results establish a versatile and scalable route for engineering flat-band photonic states, and provide new opportunities for wide-angle resonant enhancement, enhanced light-matter interaction, and polarization-selective photonic functionalities in both low- and high-index photonic systems.

## ACKNOWLEDGMENTS

This work was supported by the National Natural Science Foundation of China (Grant Nos. 92150109 and 61975018) and Beijing Key Laboratory of High-Entropy Energy materials and Devices, Beijing Institute of Nanoenergy and Nanosystems (No. GS2025ZD011).

The authors declare no conflicts of interest.

## DATA AVAILABILITY

The data that support the findings of this article are not publicly available. The data are available from the authors upon reasonable request.

## APPENDIX A: DEDAILs OF THE TWO-STEP COUPLING MODEL

In the designed PC slab, we consider four uncoupled TM-like modes of $\omega_{+1,0}^{TM}$, $\omega_{-1,0}^{TM}$, $\omega_{0,+1}^{TM}$, $\omega_{0,-1}^{TM}$ (red lines in Fig. 6(a)) and four uncoupled TE-like modes (blue lines in Fig. 6(a)) of $\omega_{+1,0}^{TE}$, $\omega_{-1,0}^{TE}$, $\omega_{0,+1}^{TE}$, $\omega_{0,-1}^{TE}$, as described by Eq. (6). The $\omega_{0,+1}^{TM}$ and $\omega_{0,-1}^{TM}$ are degenerate, and the $\omega_{0,+1}^{TE}$ and $\omega_{0,-1}^{TE}$ are degenerate. The corresponding effective group index are $n_g^{TM} = 1.56$ and $n_g^{TE} = 1.48$, respectively, and the effective index of slab is set as $n_{slab} = 1.61$ for comparison with the simulation results in Fig. 3(a). And we assume that all uncoupled modes have constant losses, the radiative losses of $\omega_{+1,0}^{TM}$, $\omega_{-1,0}^{TM}$, $\omega_{0,+1}^{TM}$, $\omega_{0,-1}^{TM}$ are $\gamma_1^{TM} = 0.261$ meV, $\gamma_2^{TM} = 0.261$ meV, $\gamma_3^{TM} = 0.653$ meV, and $\gamma_4^{TM} = 0.653$ meV, respectively. And the radiative losses of $\omega_{+1,0}^{TE}$, $\omega_{-1,0}^{TE}$, $\omega_{0,+1}^{TE}$, $\omega_{0,-1}^{TE}$ are $\gamma_1^{TE} = 2.610$ meV, $\gamma_2^{TE} = 2.610$ meV, $\gamma_3^{TE} = 3.263$ meV, and $\gamma_4^{TE} = 3.263$ meV, respectively. We now start to consider the two-step coupling mode. In the first step, intra-polarization coupling occurs among modes of the same polarization class. The resulting dispersions of coupled TM-like and TE-like modes are calculated using the Hamiltonian in Eq. (2) and shown in Fig. 6(b). The corresponding coupling constants are chosen as $U_1^{TM} = 14.031$ meV, $U_2^{TM} = 16.315$ meV, $U_3^{TM} = 22.841$ meV, $U_1^{TE} = -22.841$ meV, $U_2^{TE} = -22.841$ meV, $U_3^{TE} = -9.789$ meV. The resulting coupled TM-like modes are labeled as $\omega_{x,+}^{TM}$, $\omega_{x,-}^{TM}$, $\omega_{y,+}^{TM}$, $\omega_{y,-}^{TM}$, in descending order of frequency. Similarly, the coupled TE-like modes are labeled as $\omega_{x,+}^{TE}$, $\omega_{x,-}^{TE}$, $\omega_{y,+}^{TE}$, $\omega_{y,-}^{TE}$. In the second step, vertical mirror-symmetry breaking introduces additional inter-polarization between the coupled TM-like and TE-like modes. In particular, the symmetry-breaking-induced coupling between $\omega_{x,-}^{TE}$ and $\omega_{y,-}^{TE}$ is described by the effective Hamiltonian in Eq. (5), and the resulting dispersion is presented in Fig. 6(c). The paired flat-band modes of BIC and radiative modes formed, and the $Q$ factors of these two modes are shown in Fig. 7.

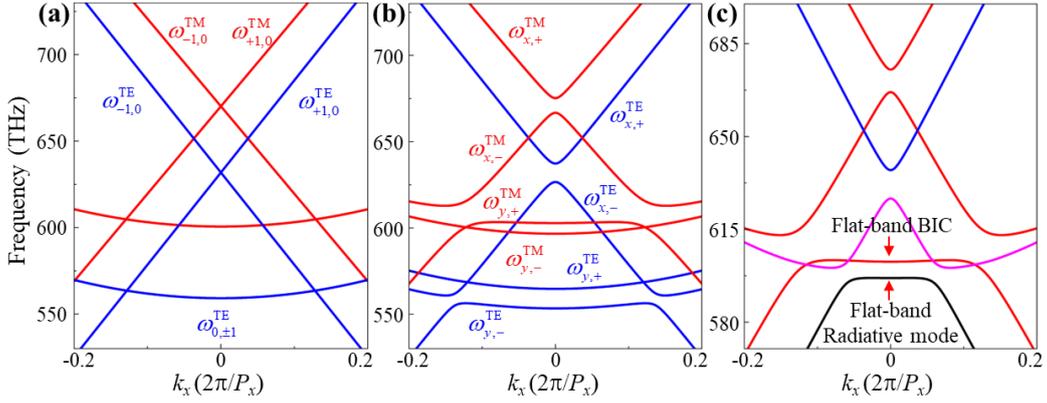

FIG. 6. (a) Uncoupled TM-like (red lines) and TE-like (blue lines) modes consider in the two-step coupling model. (b) Dispersion of coupled TM-like (red lines) and TE-like (blue lines) modes resulting from intra-polarization coupling. (c) Dispersion of hybridized induced by inter-polarization coupling .

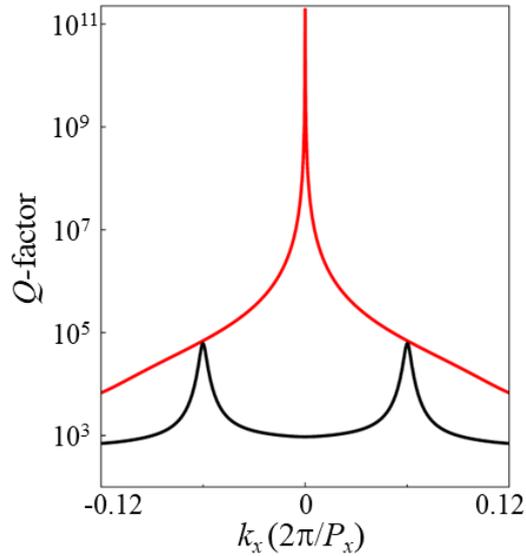

FIG. 7. $Q$ factor of the dual flat-band modes calculated from the effective Hamiltonian. The red curve corresponds to the flat-band BIC, while the black curve represents the flat-band radiative mode.

# APPENDIX B: POLARIZATION-SELECTIVE FLAT-BAND EXCITATION AND CARTESIAN MULTIPOLE DECOMPOSITION

The polarization-dependent spectral responses of the two TM-like flat-band modes are examined through angle-resolved reflection spectra calculated using the RCWA method, as shown in Fig. 8(a). The spectra are evaluated along the $x$ direction under both s- and p-polarized incidences. Under s-polarized excitation, a flat-band radiative mode is efficiently excited, whereas under p-polarized excitation, the flat-band BIC is selectively accessed.

To elucidate the physical origin of this polarization selectivity, a complete Cartesian multipole decomposition of the scattering response is performed for both resonant modes under an oblique incident angle of 4°. Oblique incidence is required because the symmetry-protected BIC cannot be directly excited at normal incidence. The total scattering power is expanded according to the generalized Cartesian multipole formalism, given by[51-53]

$$I = \frac{2\omega^4}{3c^3}|\boldsymbol{P}|^2 + \frac{2\omega^4}{3c^3}|\boldsymbol{M}|^2 + \frac{4\omega^5}{3c^4}|\boldsymbol{P}\cdot\boldsymbol{T}|^2 + \frac{2\omega^6}{3c^5}|\boldsymbol{T}|^2 + \frac{\omega^6}{5c^5}|Q_{\alpha\beta}|^2 + \frac{\omega^6}{20c^5}|M_{\alpha\beta}|^2. \quad (9)$$

Here $P_\alpha$ is electric dipole (ED) moment, $M_\alpha$ is magnetic dipole (MD) moment, $T_\alpha$ is toroidal dipole (TD) moment, $Q^E_{\alpha,\beta}$ is electric quadrupole (EQ) moment, $Q^M_{\alpha,\beta}$ is magnetic quadrupole (MQ) moment. All Cartesian multipole moments are calculated from the induced current $\boldsymbol{J}(\boldsymbol{r})$ within the unit cell of PC slab, given by:

$$P_\alpha = \frac{1}{i\omega}\int d^3 r J_\alpha(\boldsymbol{r}),$$

$$M_\alpha = \frac{1}{2c}\int d^3 r [\boldsymbol{r}\times\boldsymbol{J}(\boldsymbol{r})]_\alpha,$$

$$T_\alpha = \frac{1}{10c}\int d^3 r [(\boldsymbol{r}\cdot\boldsymbol{J}(\boldsymbol{r}))r_\alpha - 2r^2 J_\alpha(\boldsymbol{r})], \quad (10)$$

$$Q^E_{\alpha,\beta} = \frac{1}{i\omega}\int d^3 r [r_\alpha J_\beta(\boldsymbol{r}) + r_\beta J_\alpha(\boldsymbol{r}) - \frac{2}{3}\delta_{\alpha,\beta}(\boldsymbol{r}\cdot\boldsymbol{J}(\boldsymbol{r}))],$$

$$Q^M_{\alpha,\beta} = \frac{1}{3c}\int d^3 r [(\boldsymbol{r}\times\boldsymbol{J}(\boldsymbol{r}))_\alpha r_\beta + (\boldsymbol{r}\times\boldsymbol{J}(\boldsymbol{r}))_\beta r_\alpha].$$

The corresponding multipolar scattering powers are shown in Figs. 8(b) and 8(c). The flat-band radiative mode is predominantly governed by the EQ and MD contributions, whereas the quasi-BIC is mainly dominated by the TD moment and MQ components.

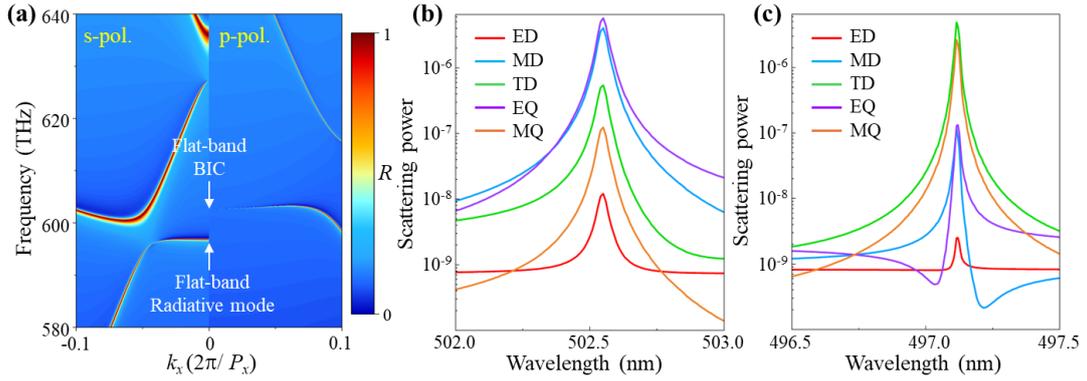

FIG. 8. (a) Angle-resolved reflection spectra along the $x$ direction under s- and p-polarized incidences. (b, c) Cartesian multipole decomposition of the scattering response of the PC slab under oblique incidence 4° for the radiative mode (b) and the quasi-BIC (c), respectively.